\documentclass[amsmat,amssymb,amsfonts,aps,prb,twocolumn]{revtex4}

\usepackage{graphicx}
\usepackage{dcolumn}
\usepackage{bm}
\usepackage{hyperref}
\usepackage{chemformula}
\usepackage{bbold}
\usepackage{amsmath}
\usepackage{siunitx}
\usepackage{comment}
\usepackage{mathtools}

\usepackage{xcolor}

\begin{document}

\title{Magneto-optical Kerr effect in spin split two-dimensional massive Dirac materials}

\author{G. Catarina$^{1,2,}$\footnote{goncalo.catarina@inl.int}, N. M. R. Peres$^{1,2}$, and J. Fern\'{a}ndez-Rossier$^{1,}$\footnote{On leave from Departamento de F\'{i}sica Aplicada, Universidad de Alicante, 03690 San Vicente del Raspeig, Spain.}}

\affiliation{$^1$QuantaLab, International Iberian Nanotechnology Laboratory (INL), 4715-330 Braga, Portugal \\
$^2$Centro de F\'{i}sica das Universidades do Minho e do Porto and Departamento de F\'{i}sica and QuantaLab, Universidade do Minho, Campus de Gualtar, 4710-057 Braga, Portugal}

\date{\today}

\begin{abstract} 
Two-dimensional (2D) massive Dirac electrons possess a finite Berry curvature, with Chern number $\pm 1/2$, that entails both a quantized dc Hall response and a subgap full-quarter Kerr rotation. 
The observation of these effects in 2D massive Dirac materials such as gapped graphene, hexagonal boron nitride or transition metal dichalcogenides (TMDs) is obscured by the fact that Dirac cones come in pairs with opposite sign Berry curvatures, leading to a vanishing Chern number. 
Here, we show that the presence of spin-orbit interactions, combined with an exchange spin splitting induced either by diluted magnetic impurities or by proximity to a ferromagnetic insulator, gives origin to a net magneto-optical Kerr effect in such systems.
We focus on the case of TMD monolayers and study the dependence of Kerr rotation on frequency and exchange spin splitting.
The role of the substrate is included in the theory and found to critically affect the results.
Our calculations indicate that state-of-the-art magneto-optical Kerr spectroscopy can detect a single magnetic impurity in diluted magnetic TMDs.
\end{abstract}

\maketitle

\section{Introduction}
The electronic states of a two-dimensional (2D) gapped Dirac Hamiltonian have a finite Berry curvature, with Chern number ${\cal C} = \pm 1/2$, that leads to~\cite{Thouless1982} a quantized dc Hall conductivity.
At finite frequencies, the Hall response is also peculiar and gives origin to a giant low-frequency Kerr rotation in thin-film topological insulators~\cite{Tse2010}.
The observation of these anomalous phenomena requires a material realization of a massive Dirac electron gas in two dimensions. 
Possible candidates are the surfaces of three-dimensional topological insulators, that host 2D massless Dirac cone states at the $\Gamma$ point of the Brillouin zone~\cite{Zhang2009}. 
Both magnetic doping~\cite{Yu2010} or spin proximity effect~\cite{Katmis2016} can be used to open up a gap, which would permit to probe the anomalous Hall response associated to massive Dirac electrons in two dimensions. 

In this work, we explore an {\em alternative route} to unveil the anomalous Hall response of 2D massive Dirac materials.
For that matter, we consider a different class of physical systems with strong light-matter coupling~\cite{Mak2010,Splendiani2010}: the widely studied~\cite{Wang2012} semiconducting transition metal dichalcogenide (TMD) monolayers, such as \ch{MoS2}.
The low-energy electronic properties of these materials are governed by states in the neighborhood of two non-equivalent valleys, which can be described in terms of a spin-valley coupled massive Dirac equation~\cite{Xiao2012}.
In the presence of time-reversal symmetry, the total Berry curvature vanishes due to a perfect cancellation of the contributions coming from the two valleys.
However, the introduction of an exchange spin splitting ---which breaks time-reversal symmetry---, combined with the strong spin-orbit interactions~\cite{Liu2013,Kosmider2013}, offsets this cancellation, leading to an anomalous Hall response~\cite{Da2017} that results in a non-vanishing magneto-optical Kerr effect.

\begin{figure}
 \includegraphics[width=\columnwidth]{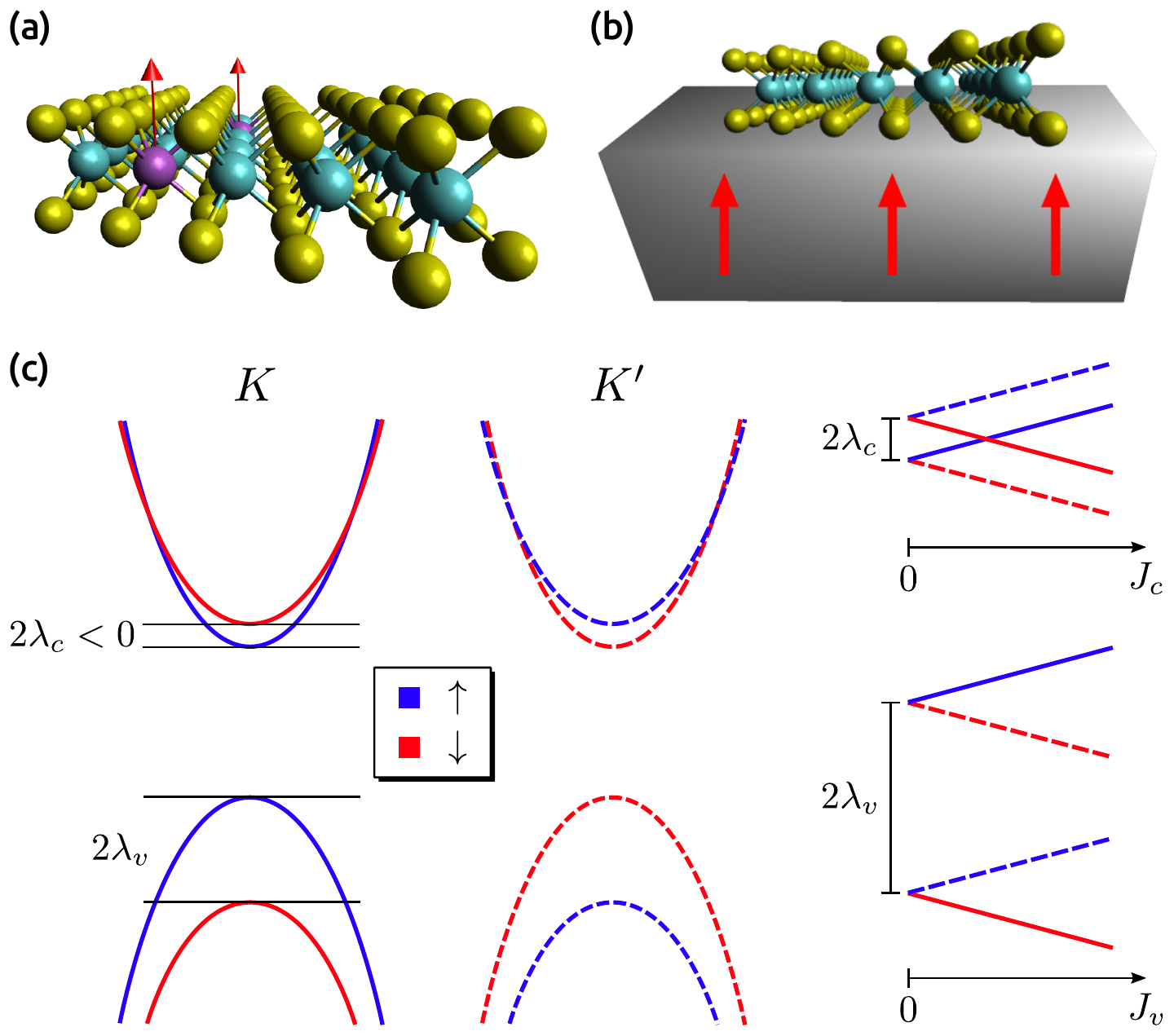}
 \caption{(Color online) (a,b) Representation of the physical system: monolayer transition metal dichalcogenide (TMD) in the presence of exchange interactions induced by diluted magnetic dopants (a) or by proximity effect to a ferromagnetic insulator (b).
 (c) Left panel: low-energy bands of monolayer TMDs.
 Blue/red lines stand for bands with spin up/down projections, split due to spin-orbit interactions with different magnitude in the conduction and valence bands, $\lambda_v \gg |\lambda_c|$.
 Solid/dashed lines represent Dirac bands obtained around the $K$/$K'$ valley.
 Right panel: effect of exchange at the bottom of the conduction bands and at the top of the valence bands.
 A combination of both spin-orbit coupling and band-dependent exchange $J_c \neq J_v$ leads to four non-degenerate effective band gaps.}
 \label{fig:system}
\end{figure}

We consider two different mechanisms to induce exchange spin splitting in TMDs.
These entail interaction of the electronic states in the valence and conduction bands of the TMD with magnetic atoms located either at the TMD itself, as magnetic dopants in a {\em diluted magnetic semiconductor} (Fig.~\ref{fig:system}a), or at an adjacent ferromagnetic insulator, in which case exchange arises from spin proximity (Fig.~\ref{fig:system}b).
Diluted magnetic doping of TMDs has been considered theoretically~\cite{Cheng2013,Ramasubramaniam2013,Mishra2013,Cheng2014,Andriotis2014} and realized experimentally~\cite{Wang2016a,Xia2017,Dau2019}.
Some intrinsic point defects in TMDs are expected to be spin polarized~\cite{Hong2015,Li2016,Khan2017,Khan2018}, so that they can also act as magnetic centers.
Exchange-driven spin splittings in TMDs caused by proximity to ferromagnetic insulators have been reported both in experiments~\cite{Zhao2017,Zhong2017,Seyler2018,Norden2019} and in first-principle calculations~\cite{Qi2015,Zhang2016,Li2018,Zollner2019}.
It has also been predicted that an antiferromagnetic layered substrate can induce, by proximity effect, a spin splitting of TMD bands~\cite{Xu2018a}; this occurs due to exchange interactions between the TMD and the surface layer of the substrate, which has ferromagnetic order.

\section{Model Hamiltonian}
We model TMD monolayers with the additional exchange spin splitting through an Hamiltonian with three terms,
\begin{equation}
\mathcal{H} = 
\mathcal{H}_{\text{MD}} + \mathcal{H}_{\text{soc}} + \mathcal{H}_{\text{ex}}.
\label{eq:H}
\end{equation}
The first two terms represent the well-known~\cite{Xiao2012} spin-valley coupled massive Dirac model for TMDs in the trigonal prismatic configuration.
These describe the low-energy electronic properties of TMDs, which are governed by states in the neighborhood of two non-equivalent points of the Brillouin zone: the so-called $K$ and $K'$ valleys.
The first term corresponds to a massive Dirac Hamiltonian,
\begin{equation}
\mathcal{H}_{\text{MD}} = 
\hbar v_{\text{F}} \left(k_x \tau_z \sigma_1 + k_y \sigma_2 \right) 
+ \frac{\Delta}{2} \sigma_3,
\end{equation}
where $\hbar$ is the reduced Planck constant, $v_{\text{F}}$ is the Fermi velocity, $\bm{k} = (k_x,k_y)$ is the electron wave vector, $\tau_z$ is the valley operator with eigenvalues $\tau=\pm$ ($+$ for $K$ and $-$ for $K'$), $\sigma_i \ (i=1,2,3)$ are Pauli matrices acting on the space of the lowest-energy conduction and highest-energy valence states, and $\Delta$ is the bare band gap (or mass, in the language of relativistic quantum mechanics).
The second term accounts for the strong spin-orbit coupling in TMDs~\cite{Liu2013,Kosmider2013}, reading as
\begin{equation}
\mathcal{H}_{\text{soc}} = 
\frac{\lambda_c (\mathbb{1} + \sigma_3) + \lambda_v (\mathbb{1} - \sigma_3)}{2} \tau_z s_z,
\end{equation}
where $2\lambda_{c/v}$ is the spin splitting in the conduction/valence bands and $s_z$ is the Pauli matrix for the out-of-plane spin component with eigenvalues $s=+(\uparrow),-(\downarrow)$.
On account of the different atomic orbital breakdown of conduction and valence states, $\lambda_v \gg |\lambda_c|$ is verified for most of the TMD materials~\cite{Liu2013}.
Importantly, spin-orbit interactions preserve time-reversal symmetry due to the so-called spin-valley coupling: states with spin $\uparrow$ in valley $K$ have a Kramers partner in valley $K'$ with spin $\downarrow$.

The third term in Eq.~\eqref{eq:H}, given by
\begin{equation}
\mathcal{H}_{\text{ex}} = 
\frac{J_c (\mathbb{1} + \sigma_3) + J_v (\mathbb{1} - \sigma_3)}{2} s_z,
\label{eq:Hex}
\end{equation}
describes an exchange-driven spin splitting of $2J_{c/v}$ in the conduction/valence bands.
This term breaks time-reversal symmetry.
In the case of diluted magnetic semiconductors, it can be derived (see Appendix~\ref{app:Kondo}) applying first-order perturbation theory to a Kondo model within the so-called virtual crystal approximation~\cite{Furdyna1988}. 
This sets
\begin{equation}
J_{c/v} = \text{x}_\text{imp} \langle M_z \rangle \gamma_{c/v}, 
\label{eq:J}
\end{equation}
where $\text{x}_\text{imp}$ is the atomic ratio of magnetic impurities, $\langle M_z \rangle$ is their statistical average spin (local spins are treated classically, within mean-field, and assumed to have orientation along $z$), and $\gamma_{c/v}$ is a material-dependent parameter, formally defined as the expectation value of the Kondo exchange coupling within conduction/valence states.
For conventional diluted magnetic semiconductors based on II-VI compounds doped with \ch{Mn}, experimental measurements yield values of $\gamma_c$ and $\gamma_v$ with opposite signs and magnitudes up to $1~\si{\electronvolt}$~\cite{Furdyna1988}.
Therefore, a {\em net exchange spin splitting} $\Delta_J = J_c - J_v$ in the order of tens of $\si{\milli\electronvolt}$ could be reached for \ch{Mn} concentrations of few percent.
In the case of TMDs on top of ferromagnetic insulators, Eq.~\eqref{eq:Hex} has been used~\cite{Qi2015,Scharf2017} to account for giant band-dependent exchange spin splittings predicted by first-principle calculations~\cite{Qi2015,Zhang2016}.

Altogether, the model Hamiltonian described by Eq.~\eqref{eq:H} can be considered as four independent copies (two per spin and valley) of a massive Dirac model, each of which with an effective gap given by
\begin{equation}
E^{\tau,s}_{\text{gap}} = 
\Delta + \tau s \Delta_\lambda + s \Delta _J,
\label{eq:eff_gap}
\end{equation}
where $\Delta_\lambda = \lambda_c - \lambda_v$.
The corresponding band spectrum is depicted in Fig.~\ref{fig:system}c.
For $J_c = J_v = 0$, TMD bands appear as two spin-valley coupled Kramers doublets, split by the strong spin-orbit interactions.
As $J_c$ and $J_v$ are ramped up, spin-valley coupling is broken.
For $\Delta_J \neq 0$, we obtain four non-degenerate effective gaps.

Zeeman splitting, in the order of $0.2~\si{\milli\electronvolt \per \tesla}$~\cite{Li2014,MacNeill2015,Aivazian2015,Srivastava2015}, is ignored in our model since, for most practical cases, it is dominated by the exchange term.

\section{Magneto-optical Kerr effect}
We are interested in the magneto-optical Kerr response of exchange spin split TMD monolayers.
Specifically, we compute the so-called Kerr rotation.
When linearly polarized light is shined into a magnetic/magnetized material, the reflected beam is in general elliptically polarized with its plane of polarization rotated with respect to that of the incident beam.
The Kerr rotation is the angle of rotation of the plane of polarization.
State-of-the-art experimental setups have reported Kerr rotation measurements with $10~\si{\nano\radian}$ resolution~\cite{Kapitulnik2009,Gong2017}.

In Appendix~\ref{app:2DMOKE}, we derive the equation that relates the complex Kerr angle and the optical conductivity of a 2D material,
\begin{equation}
\theta_{\text{K}} + \mathrm{i} \gamma_{\text{K}} \simeq 
\frac{ 2 \pi \alpha \frac{ \sigma_{xy}}{\sigma_0} }
{ \left( \pi \alpha \frac{ \sigma_{xx}}{\sigma_0} + 
\sqrt{\varepsilon_r} \right)^2 
+ \left( \pi \alpha \frac{ \sigma_{xy}}{\sigma_0} + \mathrm{i} \right)^2 }.
\label{eq:Kerr}
\end{equation}
Here, $\theta_{\text{K}}$ is the Kerr rotation, $\gamma_{\text{K}}$ is the Kerr ellipticity, $\alpha \simeq 1/137$ is the fine-structure constant, $\sigma_0 = e^2/(4\hbar)$ is the universal conductivity of graphene ($e$ is the elementary charge), $\varepsilon_r$ is the relative permittivity of the substrate on which the 2D material is deposited and $\sigma_{xx}$ ($\sigma_{xy}$) stands for the longitudinal (Hall) component of the 2D optical conductivity tensor.
The derivation of this expression assumes normal incidence (as usual within the polar geometry) of linearly polarized light onto a 2D system with $\sigma_{xx} = \sigma_{yy}$ and $\sigma_{xy} = -\sigma_{yx}$, placed on top of a semi-infinite dielectric substrate.
Moreover, it is only valid in the limit of small $\theta_{\text{K}}$ and $\gamma_{\text{K}}$ (see Eq.~\eqref{eq:MOKE_general} in Appendix~\ref{app:2DMOKE} for the general formula).
It must be noted that this equation is applicable for strictly 2D models as it depends on the 2D optical conductivity tensor, whose units are Siemens (instead of Siemens per meter).
In Appendix~\ref{app:2DMOKE_from3D}, we show that Eq.~\eqref{eq:Kerr} can be retrieved considering multiple reflections in a stratified medium made of a three-dimensional material encapsulated between air and a substrate, taking the limit $d \rightarrow 0$, where $d$ is the thickness of the material.

In order to compute the 2D optical conductivity of exchange spin split TMDs, Kubo formula is employed.
For either $\Delta_\lambda=0$ or $\Delta_J=0$, straightforward calculations yield $\sigma_{xy}=0$, which implies a vanishing magneto-optical Kerr response.
Therefore, we conclude that, considering our model Hamiltonian, magneto-optical Kerr effects are only obtained in the presence of {\em both} spin-orbit interactions and band-dependent exchange spin splittings.

In what follows, we focus on the case where the Fermi level lies inside the gap, for which the results are independent of the temperature.
In this regime, we can ignore the Landau level structure caused by an out-of-plane magnetic field given that, on its own, it does not lead to a finite Hall response~\cite{Catarina2019}.
It must be noted, however, that the single-particle description followed here does not account for the strong excitonic effects present in TMDs at charge neutrality~\cite{Wang2018}.
This subject is left for a companion publication~\cite{Henriques2020}.

Eq.~\eqref{eq:Kerr} shows that the complex Kerr angle has a non-trivial dependence both on the properties of the 2D material, encoded in its optical conductivity tensor, and on the dielectric constant of the substrate.
In this work, we consider substrates with $\varepsilon_r \gtrsim 2$, which is in principle the most natural case in experiments.
Within this assumption, we systematically find that, provided intraband transitions are Pauli blocked at charge neutrality, Eq.~\eqref{eq:Kerr} can be simplified into
\begin{equation}
\theta_{\text{K}} + \mathrm{i} \gamma_{\text{K}} \simeq 
\frac{2 \pi \alpha}{\varepsilon_r - 1} \frac{\sigma_{xy}}{\sigma_0}.
\label{eq:Kerr2}
\end{equation}
Thus, we see that Kerr rotation is governed by the real part of $\sigma_{xy}$.

The optical conductivity of our model Hamiltonian can always be expressed as a sum of spin- and valley-resolved contributions.
In the limit where Eq.~\eqref{eq:Kerr2} is valid, we can also define a spin- and valley-resolved Kerr rotation such that
\begin{equation}
\theta_{\text{K}} (\omega) =
\sum_{\tau,s} \theta^{\tau,s}_{\text{K}} (\omega),
\label{eq:Kerr_resolved}
\end{equation}
where the dependence on $\omega$, the angular frequency of the incident light, is explicitly indicated.
Using Kubo formula, we get
\begin{equation}
 \theta^{\tau,s}_{\text{K}} (\omega) =
-\tau \frac{\alpha}{\varepsilon_r - 1} 
\frac{E^{\tau,s}_{\text{gap}}}{\hbar \omega}
\log \left| \frac{\hbar \omega + E^{\tau,s}_{\text{gap}}}
{\hbar \omega - E^{\tau,s}_{\text{gap}}} \right|,
\label{eq:Kerr3}
\end{equation}
where, for simplicity, we have not included a finite empirical broadening $\Gamma$ within the Kubo formalism. 

\begin{figure*}
 \includegraphics[width=1.8\columnwidth]{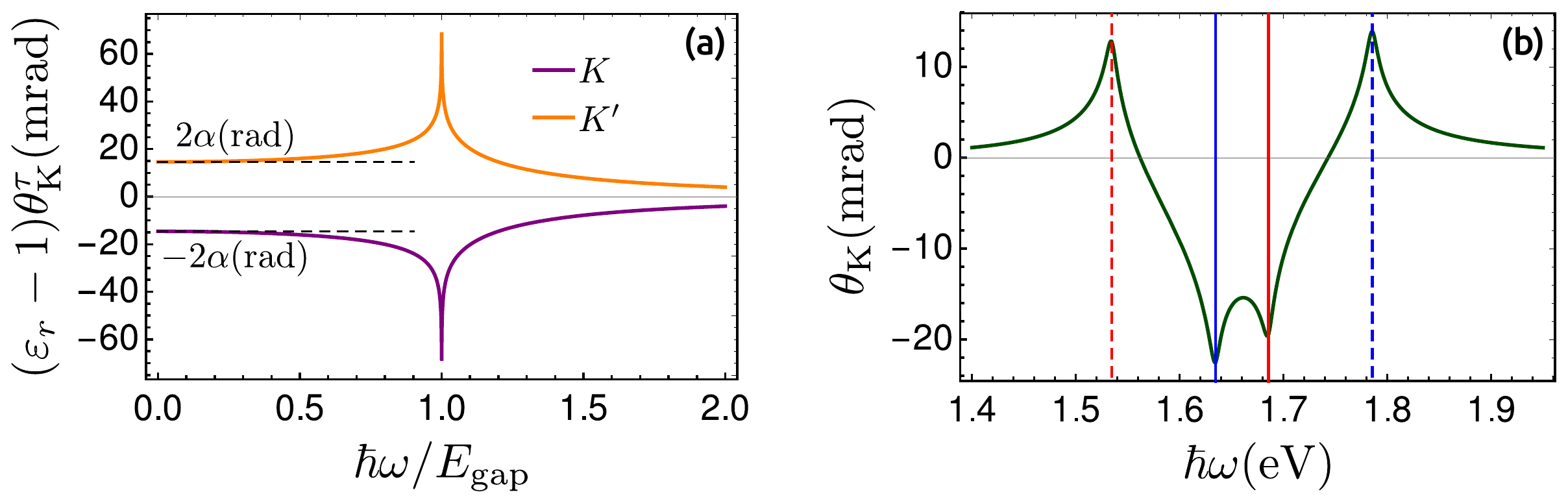}
 \caption{(Color online) (a) Valley-resolved Kerr rotation $\theta_{\text{K}}^{\tau}$, as a function of the photon energy $\hbar \omega$, for a two-dimensional massive Dirac material, with gap $E_{\text{gap}}$, at charge neutrality and placed on top of a dielectric substrate with relative permittivity $\varepsilon_r \gtrsim 2$. 
 The sum over valleys yields a vanishing net Kerr response.
 The line shape is independent of any parameter of the theory.
 The horizontal dashed lines mark low-frequency Kerr rotation plateaus at $\pm 2 \alpha / (\varepsilon_r - 1)$, where $\alpha$ is the fine-structure constant.
 (b) Kerr rotation as a function of the photon energy for monolayer \ch{MoS2} on top of \ch{SiO2} ($\varepsilon_r = 2.4$~\cite{Constant2016}), at charge neutrality and with a net exchange spin splitting $\Delta_J = 50~\si{\milli\electronvolt}$.
 Parameters: \ch{MoS2} bare band gap $\Delta = 1.66~\si{\electronvolt}$ and spin-orbit coupling splittings $2\lambda_c = -3~\si{\milli\electronvolt}$, $2\lambda_v = 148~\si{\milli\electronvolt}$~\cite{Liu2013}; empirical broadening $\Gamma = 4~\si{\milli\electronvolt}$~\cite{Ajayi2017}.
 A combination of both $\lambda_c \neq \lambda_v$ and $\Delta_J \neq 0$ leads to four non-degenerate effective band gaps ---one for each spin and valley--- that are represented by the vertical lines, following the same color and dashing style as in Fig.~\ref{fig:system}b.
 The net Kerr response can be seen as a sum of spin- and valley-resolved contributions of massive Dirac electrons (a) that are offset in energy and thus do not cancel out.}
 \label{fig:universalKerr}
\end{figure*}

The above equation describes the Kerr rotation associated to a single massive Dirac cone, with effective gap $E^{\tau,s}_{\text{gap}}$, assuming charge neutrality and a dielectric substrate with $\varepsilon_r \gtrsim 2$.
It must be noted that the line shape of $(\varepsilon_r - 1) \theta^{\tau,s}_{\text{K}} \left( \frac{\hbar \omega}{E^{\tau,s}_{\text{gap}}} \right)$ is independent of any parameter of the theory.
Taking $\Delta_\lambda = 0$ and $\Delta_J = 0$, the effective gap becomes spin- and valley-independent and, summing over $\tau$ and $s$, we obtain a vanishing Kerr rotation due to opposite sign contributions at the two valleys (Fig.~\ref{fig:universalKerr}a).
For both $\Delta_\lambda \neq 0$ and $\Delta_J \neq 0$, the presence of four non-degenerate effective gaps offsets this cancellation, leading to a {\em net Kerr rotation}, as we show in Fig.~\ref{fig:universalKerr}b.

In the dc limit, Eq.~\eqref{eq:Kerr3} gives
\begin{equation}
 \theta^{\tau}_{\text{K}} (\omega \rightarrow 0) =
-\tau \frac{2 \alpha}{\varepsilon_r - 1}.
\end{equation}
For frequencies below the gap, Eq.~\eqref{eq:Kerr3} approaches the dc limit rapidly, leading to nearly flat low-frequency plateaus, as shown in Fig.~\ref{fig:universalKerr}a.
As a side note, we stress that, if we take $\varepsilon_r = 1$ and consider a model Hamiltonian with a single massive Dirac cone, Eq.~\eqref{eq:Kerr} is no longer valid, as the corresponding Kerr rotation is not small.
Indeed, using the general formula derived in Appendix~\ref{app:2DMOKE} (Eq.~\eqref{eq:MOKE_general}), we obtain subgap full-quarter plateaus, $\theta^{\tau}_{\text{K}} (\omega < E_{\text{gap}}) \simeq -\tau \pi / 2$.
Thus, we see that the inclusion of a substrate in the theory can significantly affect the results.

We now address the properties of the (net) Kerr rotation in exchange spin split TMDs.
Taking the low-frequency limit of Eqs.~\eqref{eq:Kerr_resolved} and \eqref{eq:Kerr3}, we obtain
\begin{equation}
\theta_{\text{K}} (\omega \ll \Delta) \simeq 
- \frac{16 \alpha}{\varepsilon_r - 1} 
\left( \frac{\hbar \omega}{\Delta} \right)^2 
\frac{\Delta_J \Delta_\lambda}{\Delta^2},
\label{eq:net_Kerr_subgap}
\end{equation}
where we have also assumed $\Delta \gg \Delta_J, \Delta_\lambda$.
At higher frequencies, no simple analytical expression can be found.
This can be understood by the fact that the Kerr rotation is the sum of four curves (one per spin and valley) that have a resonant peak at the absorption thresholds defined by the corresponding effective gap, as shown in Fig.~\ref{fig:universalKerr}b.

\begin{figure*}
 \includegraphics[width=1.8\columnwidth]{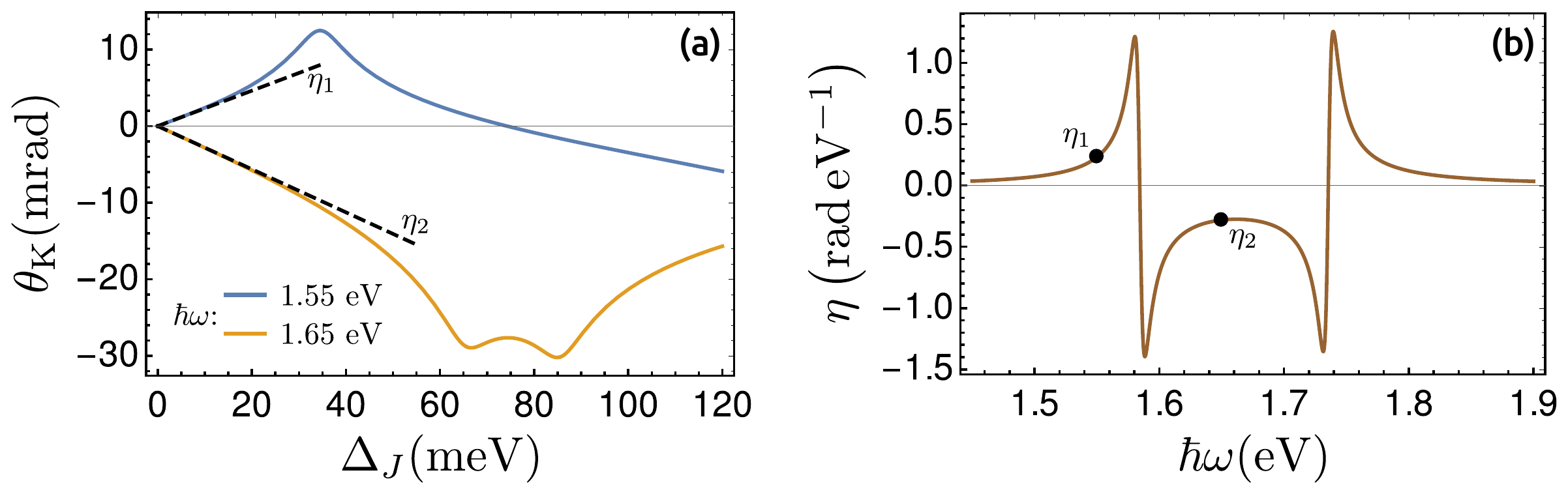}
 \caption{(Color online) (a) Kerr rotation, as a function of the net exchange spin splitting, for monolayer \ch{MoS2} on top of \ch{SiO2}, at charge neutrality, for two photon energies.
 Parameters as in Fig.~\ref{fig:universalKerr}b.
 Linear dependence is observed in the small $\Delta_J$ limit; the corresponding slopes, marked by the dashed lines, are verified to depend strongly on $\hbar \omega$.
 For larger $\Delta_J$, a non-monotonous dependence is obtained.
 The origin of this behavior is the shifting of absorption thresholds (as the ones marked by the vertical lines in Fig.~\ref{fig:universalKerr}b) as $\Delta_J$ is ramped up, which causes resonant peaks to cross the photon energy.
 (b) Derivative of Kerr rotation with respect to net exchange spin splitting $\Delta_J$, evaluated at $\Delta_J = 0$, as a function of the photon energy, for the same system as in (a).
 Marked points correspond to the slopes shown in (a).}
 \label{fig:Kerr_exchange}
\end{figure*}

Eq.~\eqref{eq:net_Kerr_subgap} shows that, in the low-frequency regime, Kerr rotation varies linearly with $\Delta_J$ and thereby with the average magnetization of the impurities (in the case of diluted magnetic TMD semiconductors), by virtue of Eq.~\eqref{eq:J}.
For frequencies close to the absorption thresholds, the linearity breaks above a given value of $\Delta_J$, as we show in Fig.~\ref{fig:Kerr_exchange}a.
This is explained by the fact that, as we vary $\Delta_J$, absorption thresholds are shifted in a way that they eventually cross the photon energy, leading to a non-monotonous dependence.
Within the linear regime, it is also evident from Fig.~\ref{fig:Kerr_exchange}a that the slopes depend strongly on $\omega$.
The frequency dependence of these slopes, defined as
\begin{equation}
\eta (\omega) = 
\left. 
\frac{\partial \theta_{\text{K}} (\omega)}{\partial \Delta_J} 
\right\rvert_{\Delta_J = 0},
\end{equation}
is presented in Fig.~\ref{fig:Kerr_exchange}b.

Finally, we focus on the case of diluted magnetic TMD semiconductors and estimate the limits of magnetic moment detection through Kerr rotation measurements.
Specifically, given an experimental setup that permits to detect Kerr rotation with resolution $\theta^{\text{res}}_{\text{K}}$, we address the question of what is the smallest number of impurities that can be probed.
We assume that we are in the regime where Kerr rotation scales linearly with exchange, $\theta_{\text{K}} (\omega) = \eta(\omega) \Delta_J$.
Using Eq.~\eqref{eq:J}, we can thus write
\begin{equation}
\left| 
\eta(\omega) \text{x}_\text{imp} \langle M_z \rangle 
( \gamma_c - \gamma_v ) 
\right| 
> \theta^{\text{res}}_{\text{K}}.
\end{equation}
Taking the Abbe diffraction limit, we consider a laser spot with area $A_{\text{spot}} = \pi \left( \frac{\lambda}{4 \text{NA}} \right)^2$, where $\lambda$ is the wave length of the light and NA is the numerical aperture of the laser.
Assuming a maximum of one impurity per unit cell, the number of impurities probed by the laser spot can be written as $N_{\text{imp}} = \text{x}_\text{imp} \frac{A_{\text{spot}}}{A_{\text{u.c.}}}$, where $A_{\text{u.c.}} = \frac{\sqrt{3}}{2} a^2$ is the area of the unit cell ($a$ is the lattice parameter).
With this, we get
\begin{equation}
N_{\text{imp}} > 
\frac{\pi \lambda^2}{6 a^2 \text{NA}^2}
\frac{\theta^{\text{res}}_{\text{K}}}
{\left| \eta(\omega) \langle M_z \rangle ( \gamma_c - \gamma_v ) \right|}.
\end{equation}
In order to give rough estimations, we take $a = 3.2~\si{\angstrom}$~\cite{Xiao2012} (having \ch{MoS2} as reference), $\text{NA} \sim 1$, $\theta^{\text{res}}_{\text{K}} = 10~\si{\nano\radian}$~\cite{Kapitulnik2009,Gong2017}, $\langle M_z \rangle \sim 1$ and $\left| \gamma_c - \gamma_v \right| \sim 1~\si{\electronvolt}$~\cite{Furdyna1988} (taking as reference conventional diluted magnetic semiconductors).
For low frequencies, $\eta(\omega)$ can be obtained analytically through Eq.~\eqref{eq:net_Kerr_subgap}.
Replacing $\Delta = 1.66~\si{\electronvolt}$~\cite{Liu2013}, $\Delta_\lambda = -151~\si{\milli\electronvolt}$~\cite{Liu2013} and $\varepsilon_r = 3.9$~\cite{Constant2016} (considering \ch{MoS2} on top of \ch{SiO2} and taking the dc limit of its relative permittivity), we get $N_{\text{imp}} \gtrsim \frac{100}{\left( \hbar \omega [ \si{\electronvolt} ] \right)^2}$.
At higher frequencies, we use the results of Fig.~\ref{fig:Kerr_exchange}b to obtain $\eta(\omega)$.
Following a conservative approach, we avoid resonances and set $\hbar \omega = 1.65~\si{\electronvolt}$, for which $\eta(\omega)$ takes the value marked as $\eta_2$.
This leads to $N_{\text{imp}} \gtrsim 0.1$, showing that a single impurity can be detected.
It must be noted that magneto-optical effects have been used for single spin detection~\cite{Berezovsky2006,Atature2007}.

\section{Discussion and conclusions}
We have presented a theory for magneto-optical Kerr effects in 2D materials whose low-energy bands are described by a spin split massive Dirac equation.
Using the standard Fresnel formalism, we have obtained the equation that relates the complex Kerr angle with the optical conductivity tensor of a 2D system, considering the effect of a substrate.
We have found that a combination of both spin-orbit interactions and band-dependent spin splitting in the model leads to an anomalous Hall conductivity that gives origin to a non-vanishing magneto-optical Kerr response.
We have focused our theory in transition metal dichalcogenide monolayers, for which spin-orbit interactions are strong, and considered an exchange spin splitting induced either by diluted magnetic impurities or by proximity effects to a ferromagnetic insulator.
Our formalism can be extended to tight-binding Hamiltonians and to other types of magnetic order~\cite{Sivadas2016}.

The main results, obtained at charge neutrality and for substrates with relative permittivity $\varepsilon_r \gtrsim 2$, are the following.
First, we have obtained a simplified expression which shows that Kerr rotation is governed by the real part of the Hall conductivity and therefore permits to define spin- and valley-resolved contributions.
Second, we have shown that a single valley of a 2D gapped Dirac model (with gap $E_{\text{gap}}$) entails a Kerr rotation with opposite sign for each of the valleys and whose frequency dependence is given by a function that depends only on $\omega/E_{\text{gap}}$, taking the value $-\tau \frac{2 \alpha}{\varepsilon_r - 1}$ in the dc limit, where $\tau=\pm$ is the valley index and $\alpha$ is the fine-structure constant.
Third, we have seen that the model Hamiltonian for exchange spin split TMDs can be considered as four copies (two per spin and valley) of a gapped Dirac equation with non-degenerate effective gaps, such that the (net) Kerr rotation can be interpreted as a non-cancellation of spin- and valley-resolved features of a 2D massive Dirac theory.
Fourth, we have addressed the use of Kerr rotation measurements to probe magnetic moments in diluted magnetic TMD semiconductors, showing that state-of-the-art experimental setups can detect signal coming from a single impurity.

The role of excitonic corrections will be the subject of future work.

\section*{Acknowledgments}
We thank Allan H. MacDonald, Elaine Li, Alejandro Molina-S\'anchez and Jo\~{a}o C. G. Henriques for fruitful discussions.
G. C. acknowledges Funda\c{c}\~{a}o para a Ci\^{e}ncia e a Tecnologia (FCT) for Grant No. SFRH/BD/138806/2018.
G. C. and J. F.-R. acknowledge financial support from FCT through Grant No. P2020-PTDC/FIS-NAN/4662/2014.
N. M. R. P. acknowledges financial support from European Commission through project ``Graphene-Driven Revolutions in ICT and Beyond'' (Ref. No. 785219), FCT in the framework of Strategic Financing (Ref. No. UID/FIS/04650/2019), and COMPETE2020, PORTUGAL2020, FEDER and FCT for Grants No. PTDC/FIS-NAN/3668/2013, No. POCI-01-0145-FEDER-028114, No. POCI-01-0145-FEDER-029265 and No. PTDC/NAN-OPT/29265/2017.
J. F.-R. acknowledges FCT for Grant No. UTAP-EXPL/NTec/0046/2017, as well as Generalitat Valenciana funding Prometeo2017/139 and MINECO-Spain (Grant No. MAT2016-78625-C2).

\appendix

\section{Exchange spin splitting in diluted magnetic semiconductors} 
\label{app:Kondo}
Following Ref.~\onlinecite{Furdyna1988}, we model the exchange interaction between band electrons and diluted magnetic impurities through a Kondo-like exchange term,
\begin{equation}
 \mathcal{V} = 
 \sum_{\bm{R}_i} J(\bm{R}_i) \bm{M}(\bm{R}_i) \cdot \bm{s}, 
\end{equation}
where $\bm{M}(\bm{R}_i)$ is the vector of Pauli operators for the spin of magnetic impurities located at positions $\bm{R}_i$, $\bm{s}$ is the vector of Pauli operators for the spin of band electrons and $J(\bm{R}_i)$ are exchange coupling constants.

Treating the local spins classically and within mean-field, we replace $\bm{M}(\bm{R}_i)$ by its statistical average $\langle \bm{M} \rangle$.
In addition, we assume an average magnetization along the $z$ direction.
Moreover, we employ the so-called virtual crystal approximation, making $\sum_{\bm{R}_i} J(\bm{R}_i) \rightarrow \text{x}_\text{imp} \sum_{\bm{R}} J(\bm{R})$, where $\text{x}_\text{imp}$ is the atomic ratio of magnetic impurities and $\bm{R}$ denotes the positions of lattice sites.
With this, we get
\begin{equation}
 \mathcal{V} = 
 s_z \langle M_z \rangle \text{x}_\text{imp} \sum_{\bm{R}} J(\bm{R}). 
 \label{eq:V_vca}
\end{equation}

To first order in perturbation theory, Eq.~\eqref{eq:V_vca} leads to a correction of the energy levels given by
\begin{equation}
 \delta E^{(1)} = 
 s \langle M_z \rangle \text{x}_\text{imp} 
 \langle \psi_0 | \sum_{\bm{R}} J(\bm{R}) | \psi_0 \rangle,
\end{equation}
where $\psi_0$ is the wave function of the unperturbed Hamiltonian, which is assumed to be diagonal in the subspace of $s_z$ with eigenvalues $s=+(\uparrow),-(\downarrow)$.
We now notice that the matrix element present in the above equation depends on the atomic orbital breakdown of $\psi_0$, such that it can be different for electrons in distinct bands.
Taking this into account, and considering the low-energy Dirac model for TMD monolayers, we write
\begin{equation}
 \delta E^{(1)}_{c/v} = 
 s \langle M_z \rangle \text{x}_\text{imp} \gamma_{c/v}
\end{equation}
where $\gamma_{c/v}$ is the matrix element of the exchange coupling constants for conduction/valence states.
Finally, we define 
\begin{equation}
 J_{c/v} = 
 \langle M_z \rangle \text{x}_\text{imp} \gamma_{c/v},
\end{equation}
such that $2J_{c/v}$ is the exchange spin splitting in the conduction/valence bands, as captured by Eq.~\eqref{eq:Hex}.

\section{Magneto-optical Kerr effect in two-dimensional systems}

\subsection{Formalism}
\label{app:2DMOKE}
We consider a 2D system lying at the $xy$ plane, with air above ($z>0$) and a substrate below ($z<0$). 
We treat air as vacuum and assume a non-magnetic dielectric substrate with relative permittivity $\varepsilon_{r}$. 
Both media are taken as semi-infinite, disregarding any phenomenon of multiple reflections in stratified media.

We assume normal incidence of linearly polarized monochromatic light and write its electric field as
\begin{equation}
 \bm{E}^{\text{(i)}} (z,t) = 
 E_x^{\text{(i)}} \bm{u}_x 
 \mathrm{e}^{\mathrm{i} \left(-\frac{\omega}{c} z - \omega t \right)},
\end{equation}
where $t$ is the time, $\omega$ is the angular frequency of the light and $c$ is the speed of light in vacuum.
The electric field of the reflected and the transmitted light can be written as
\begin{equation}
 \bm{E}^{\text{(r)}} (z,t) = 
 \left( E_x^{\text{(r)}} \bm{u}_x + E_y^{\text{(r)}} \bm{u}_y \right)
 \mathrm{e}^{\mathrm{i} \left(\frac{\omega}{c} z - \omega t \right)}
\end{equation}
and
\begin{equation}
 \bm{E}^{\text{(t)}} (z,t) = 
 \left( E_x^{\text{(t)}} \bm{u}_x + E_y^{\text{(t)}} \bm{u}_y \right)
 \mathrm{e}^{\mathrm{i} 
 \left(-\sqrt{\varepsilon_r} \frac{\omega}{c} z - \omega t \right)},
\end{equation}
respectively.
The corresponding magnetic fields $\bm{B}^{\text{(i)}}$, $\bm{B}^{\text{(r)}}$ and $\bm{B}^{\text{(t)}}$ are obtained via Maxwell's equations.

The interface conditions at $z=0$ impose
\begin{equation}
 \bm{u}_z \times 
 \left[ 
 \bm{E}^{\text{(i)}} (0,t) + \bm{E}^{\text{(r)}} (0,t) 
 - \bm{E}^{\text{(t)}} (0,t) 
 \right] 
 = \bm{0},
\end{equation}
\begin{equation}
 \bm{u}_z \times 
 \left[ 
 \bm{B}^{\text{(i)}} (0,t) + \bm{B}^{\text{(r)}} (0,t) 
 - \bm{B}^{\text{(t)}} (0,t) \right] 
 = \mu_0 \bm{j}_s,
\end{equation}
where $\mu_0$ is the vacuum permittivity and $\bm{j}_s$ is the surface current density at the $z=0$ plane.
Applying Ohm's law, we write
\begin{equation}
 \bm{j}_s = 
 \begin{pmatrix}
  \sigma_{xx} & \sigma_{xy} \\
  \sigma_{yx} & \sigma_{yy}
 \end{pmatrix} 
 \cdot  
 \begin{pmatrix}
  E_x^{\text{(t)}} (0,t) \\
  E_y^{\text{(t)}} (0,t) 
 \end{pmatrix},
\end{equation}
where $\sigma_{ab} \ (a,b=x,y)$ are the components of the optical conductivity tensor of the 2D material.
Assuming that $\sigma_{xx} = \sigma_{yy}$ and $\sigma_{xy} = - \sigma_{yx}$, straightforward manipulation permits to obtain the reflection coefficients for right/left-handed light as
\begin{equation}
 r_\pm = 
 \frac{E_\pm^{\text{(r)}}}{E_x^{\text{(i)}} / \sqrt{2}} = 
 \frac{1 - \sqrt{\varepsilon_r} - c \mu_0 \sigma_\mp}
 {1 + \sqrt{\varepsilon_r} + c \mu_0 \sigma_\mp},
 \label{eq:r_pm}
\end{equation}
where $\sqrt{2} E_\pm^{\text{(r)}} = E_x^{\text{(r)}} \pm \mathrm{i} E_y^{\text{(r)}}$ and $\sigma_\pm = \sigma_{xx} \pm \mathrm{i} \sigma_{xy}$.

On the other hand, we can also write
\begin{equation}
 \frac{r_+}{r_-} = 
 \frac{\left| E_+^{\text{(r)}} \right|}{\left| E_-^{\text{(r)}} \right|}
 \mathrm{e}^{\mathrm{i} \left( \phi_+ - \phi_- \right)},
\end{equation}
where $E_\pm^{\text{(r)}} = \left| E_\pm^{\text{(r)}} \right| \mathrm{e}^{\mathrm{i} \phi_\pm}$.
We now identify the Kerr rotation as
\begin{equation}
 \theta_{\text{K}} = 
 \frac{\phi_- - \phi_+}{2}
\end{equation}
and the Kerr ellipticity $\gamma_{\text{K}}$ through
\begin{equation}
 \tan{\gamma_{\text{K}}} = 
 \frac{\left| E_+^{\text{(r)}} \right| - \left| E_-^{\text{(r)}} \right|}
 {\left| E_+^{\text{(r)}} \right| + \left| E_-^{\text{(r)}} \right|}.
\end{equation}
As final result, we get
\begin{widetext}
\begin{equation}
 \tan{\left( \gamma_{\text{K}} + \frac{\pi}{4} \right)} 
 \mathrm{e}^{-\mathrm{i} 2\theta_{\text{K}}} = 
 \left( \frac{1 - \sqrt{\varepsilon_r} - c \mu_0 \sigma_-}
 {1 + \sqrt{\varepsilon_r} + c \mu_0 \sigma_-} \right)
 \left( \frac{1 + \sqrt{\varepsilon_r} + c \mu_0 \sigma_+}
 {1 - \sqrt{\varepsilon_r} - c \mu_0 \sigma_+} \right),
 \label{eq:MOKE_general}
\end{equation}
\end{widetext}
which is the general formula for magneto-optical Kerr effect in 2D systems.
The above equation shows that a non-vanishing magneto-optical Kerr response implies $\sigma_+ \neq \sigma_-$, which in turn implies $\sigma_{xy} \neq 0$.

In the limit of small $\theta_{\text{K}}$ and $\gamma_{\text{K}}$, we write $\tan{\left( \gamma_{\text{K}} + \frac{\pi}{4} \right)} \mathrm{e}^{-\mathrm{i} 2\theta_{\text{K}}} \simeq 1 - \mathrm{i} 2\theta_{\text{K}} + 2\gamma_{\text{K}}$ and Eq.~\eqref{eq:MOKE_general} is simplified into Eq.~\eqref{eq:Kerr}.

\subsection{Agreement with the three-dimensional case}
\label{app:2DMOKE_from3D} 
We now consider a stratified medium made of a magnetic material with thickness $d$, encapsulated between air ($z>0$) and a substrate ($z < -d$).
As in Section~\ref{app:2DMOKE}, we treat air as vacuum and assume a non-magnetic dielectric substrate with relative permittivity $\varepsilon_{r}$. 
Both air and the substrate are taken as semi-infinite but, in contrast to the previous derivation, the finite thickness of the magnetic medium obliges us to account for multiple reflections within the Fresnel formalism. 
Regarding the properties of the magnetic material, we assume that its permittivy tensor can be expressed as
\begin{equation}
 \varepsilon^\text{MM} = \varepsilon_0
 \begin{pmatrix}
  \varepsilon_{xx} &  \varepsilon_{xy} & 0 \\
  -\varepsilon_{xy} & \varepsilon_{xx} & 0 \\
  0 & 0 & \varepsilon_{zz}
 \end{pmatrix},
\end{equation}
where $\varepsilon_0$ is the vacuum permittivity.

Similarly to what was done in the previous section, we obtain the reflection coefficients by imposing the interface conditions at $z=0$ and $z=-d$.
The major difference is that, in the magnetic medium, the propagation of the light is not isotropic. 
Indeed, the electric field of the light has components along the $\bm{u}_{\pm} = \left(\bm{u}_{x} \pm \mathrm{i} \bm{u}_{y} \right) / \sqrt{2}$ directions with (complex) refractive indexes given by $n_\pm=\sqrt{\varepsilon_{xx}\pm\text{i}\varepsilon_{xy}}$.
In addition, surface currents are now disregarded.
After some straightforward algebra, we get
\begin{equation}
 r_\pm = 
 \frac{1 - h(n_\mp)}{1 + h(n_\mp)},
 \label{eq:r_pm_3D}
\end{equation}
with
\begin{equation}
 h (n_\pm) = 
 n_\pm \frac{f\left( n_\pm \right) - g\left( n_\pm \right)}
 {f\left( n_\pm \right) + g\left( n_\pm \right)},
\end{equation}
\begin{equation}
 f\left( n_\pm \right) = 
 \left( n_\pm + \sqrt{\varepsilon_r} \right)
 \mathrm{e}^{-\mathrm{i} \frac{\omega}{c} n_\pm d},
\end{equation}
\begin{equation}
 g\left( n_\pm \right) = 
 \left( n_\pm - \sqrt{\varepsilon_r} \right)
 \mathrm{e}^{\mathrm{i} \frac{\omega}{c} n_\pm d},
\end{equation}
which is the general formula in the three-dimensional case.

In the limit of $d\rightarrow0$, we relate the optical conductivity in two and three dimensions via
\begin{equation}
 \sigma^{\text{3D}} =
 \frac{\sigma^{\text{2D}}}{d}.
\end{equation}
Using the general (tensor) relation
\begin{equation}
 \varepsilon = 
 \varepsilon_0 \mathbb{1} + \mathrm{i} \frac{\sigma^{\text{3D}}}{\omega}
\end{equation}
and expanding equation Eq.~\eqref{eq:r_pm_3D} to leading order in $d$, we obtain
\begin{equation}
 r_\pm = 
 \frac{1 - \sqrt{\varepsilon_r} - c \mu_0 \sigma^{\text{2D}}_\mp}
 {1 + \sqrt{\varepsilon_r} + c \mu_0 \sigma^{\text{2D}}_\mp},
\end{equation}
thus recovering Eq.~\eqref{eq:r_pm}.

\bibliographystyle{apsrev4-1}

\end{document}